\documentclass[runningheads]{llncs}

\usepackage{pgfplots}
\usepgfplotslibrary{groupplots}
\usetikzlibrary{decorations.pathmorphing}
\pgfplotsset{compat=1.18}
\usepackage{amssymb}
\usetikzlibrary{patterns}

\usepackage[T1]{fontenc}
\usepackage{graphicx}
\usepackage{listings}
\usepackage{booktabs} 
\usepackage{url}
\usepackage[inline,shortlabels]{enumitem}
\usepackage{tcolorbox}
\usepackage{todonotes}
\usepackage{wasysym}
\usepackage{changepage}
\usepackage{wrapfig}
\usepackage{csquotes}
\usepackage{fmtcount}
\usepackage{xspace}

\definecolor{OliveGreen}{cmyk}{0.64,0,0.95,0.40}

\newcommand{\Figure}[1]{Figure~\ref{#1}}
\newcommand{\Table}[1]{Table~\ref{#1}}

\newcommand{\HIDE}[1]{}

\newcommand{\RQ}[1]{\textbf{RQ\textsubscript{#1}}}

\newenvironment{fancyquote*}[1][]{\def\attribution{#1}``\unskip\small\itshape\ignorespaces}{\unskip{''} \mbox{\textsc{\em\scriptsize\attribution}}}

\newcommand{\Cpp}{\mbox{C\texttt{++}}}

\begin{document}
\title{Orchestrating AI-Assisted Code Remediation:\\ Socio-Technical Bottlenecks in a\\ Large Industrial Repository}
\titlerunning{Orchestrating AI-Assisted Code Remediation}


%
%
\author{Andreas Bexell\inst{1,2}\orcidID{0009-0003-1356-303X} \and
\\Lo Gullstrand Heander\inst{2}\orcidID{0000-0002-0695-4580} \and \\
Emma Söderberg\inst{2}\orcidID{0000-0001-7966-4560} \and Sigrid Eldh\inst{1,3}\orcidID{0000-0002-5070-9312}}

\authorrunning{Bexell et al.}
%
\institute{Ericsson AB, Sweden \and
Lund University, Sweden \and Mälardalen University, Sweden \& Carleton University, Canada\\
\email{\{andreas.bexell, sigrid.eldh\}@ericsson.com}\\
\email{\{lo.gullstrand\_heander, emma.soderbergh\}@cs.lth.se}\\
}
\maketitle              
\begin{abstract}
\emph{Background:} Code degradation in large, long-lived codebases is costly to remediate through manual refactoring and opportunistic clean-ups. 
LLM-based coding assistants can perform mechanical remediation at scale, but their impact on industrial workflows is underexplored.
\emph{Objective:} We investigate how massive AI-assisted code remediation affects build-on-commit continuous integration (CI), code review, and team coordination in a large industrial repository, and which socio-technical bottlenecks constrain such remediation when source editing becomes cheap through AI assistance.
\emph{Method:} We report on a 15-day exploratory single-case field study in which an experienced developer used a command-line AI coding buddy to remediate widespread issues in a closed-source industrial \Cpp{} repository. We triangulate Gerrit metadata with a developer diary and team chat, analyzed through descriptive statistics and qualitative coding.
\emph{Results:} AI-assisted remediation rapidly generated hundreds of commits touching thousands of lines, saturating CI and reviewer attention. Naïve per-file commits overloaded build-on-commit CI; Switching to directory-based batching and capping the number of files per change restored throughput, but still required explicit review solicitation, negotiation of acceptable commit granularity, and iterative follow-up to resolve build and static-analysis failures.
\emph{Conclusion:} When mechanical editing is cheap, CI capacity, review effort, and change orchestration become primary bottlenecks. Sustainable AI-assisted remediation in very large repositories requires deliberate control of commit, review, and CI batch granularity and treating semantic change sets, such as ``fix all instances of warning X'', as first-class units of work that can be sliced differently for developers, reviewers, and CI.

\end{abstract}

\section{Introduction}

Over time, code bases tend to succumb to code degradation~\cite{Jaspan2023TechnicalDebt}, a kind of technical debt where code has not kept up with evolving standards and is in need of refactoring or updates.
Static analysis can prevent the proliferation of new degradation, but it typically does not repair existing issues. In practice, improvement is often left to the ``Boy Scout Rule'': leave the code cleaner than you found it, which is slow and difficult to distinguish from feature work \cite{martin2009clean,ciolkowski2017lessons}.

The recent shift towards AI-driven software development is adding further reasons to strive to prevent code degradation.
Recent work on ``AI-friendly'' code shows that higher code health is associated with substantially lower break rates when LLMs refactor existing code, while low-quality code may be unfit for reliable AI intervention\cite{borg2026codeForMachines}.
Investing in maintainability thus benefits both human comprehension and safe, effective use of AI assistants.

Beyond opportunistic clean-ups, systematic remediation has long been possible with custom tools and scripts, such as Coccinelle~\cite{lawall2010finding} and example-based systematic editing~\cite{jacobellis2013lase}. However, their use requires specialized knowledge of these systems.
Modern AI coding assistants now enable developers to perform such mechanical remediation efficiently and at scale without specialized scripts and knowledge.
This raises the following research questions:

\RQ{1} How does massive AI-assisted code remediation affect continuous integration (CI), code review practices, and team coordination in a large industrial repository?

\RQ{2} When source editing becomes cheap through AI assistance, what socio-technical bottlenecks constrain massive code remediation in a large industrial repository?

To address these questions, we conducted a 15-day exploratory study in the context of a large international company with industrial code development.
The codebase has been maintained and expanded for decades, contains hundreds of thousands of commits, and comprises more than 10 million lines of code in six languages, the dominant one being \Cpp.
A developer with substantial software engineering experience, but limited prior exposure to contemporary AI coding assistants, used a command-line, chat-based assistant to discover and address technical debt and subsequently uploaded the AI-assisted changes to the existing CI and code-review pipeline.
We collected data through a diary, the developer's online interactions with their team, log from the coding assistant, and metadata from the code review system.

We contribute an in-situ industrial case study of AI-assisted remediation in a long-lived, CI-intensive product, showing how large-scale mechanical clean-ups shift the main bottlenecks from editing effort to CI capacity, review attention, and change orchestration. We provide practical strategies for batching and workflow design that make such AI-assisted improvement sustainable in everyday development.
\section{Related Work}
\label{sec:related}

\paragraph{Technical debt, code degradation, and refactoring practice.}
Code degradation and related forms of technical debt have long been recognized as drivers of increased maintenance cost and reduced development velocity.
Refactoring is often proposed as the primary means to counteract such degradation, but empirical studies show that refactoring is itself perceived as costly and risky work, as found by Kim et al.~\cite{kim2012field}
In a large field study at Microsoft, they combined a survey of over 300 engineers with an analysis of Windows~7 version history.
They found that practitioners' notion of refactoring extends beyond textbook, behavior-preserving transformations and typically involves larger efforts to improve readability, maintainability, and performance, frequently interleaved with feature and bug-fix work. They also report that centrally planned refactorings, supported by custom tools and dependency analysis, correlate with reduced inter-module dependencies and fewer post-release defects.
Murphy-Hill and Black distinguish \emph{floss refactoring}: small refactorings interleaved with feature or bug-fix development; from \emph{root-canal refactoring}: focused, dedicated refactoring work~\cite{murphy2008refactoring}.
We adopt \emph{floss refactoring} to denote such small-scale clean-ups.
Our work complements these findings by exploring what happens when the \emph{mechanical} cost of such clean-ups is drastically reduced by an AI assistant and by focusing on the new bottlenecks that emerge in CI, code review, and team coordination when many small clean-ups are proposed at once.

\paragraph{Large-scale refactoring in industry.}
Ivers et al. studied large-scale refactoring (LSR) through a survey of 107 practitioners from multiple organizations~\cite{ivers2022industry}.
They define LSR as restructuring software without adding new functionality, to improve non-functional qualities or change the architecture, typically over months to years, and with substantial resource investments on systems of 100k--1M+ lines of code.
Their results show that LSR is common and that many systems experience multiple LSRs during their lifetime, but that these efforts are often postponed because of opportunity cost. New features and short-term delivery goals tend to be prioritized, even though postponing refactoring leads to slower feature delivery and deteriorating internal and external quality over time~\cite{ivers2022industry}.
In contrast to this architecture-level notion of scale, our study examines a short (15-day) exploration in which a single developer, assisted by an AI coding assistant, proposes thousands of small, largely mechanical changes in a codebase of roughly 10 million lines.
Here, ``scale'' comes from the \emph{volume and speed} of local clean-ups rather than from the duration or architectural scope of the refactoring program.
We thus focus on the socio-technical consequences of massive remediation at the level of CI capacity, review effort, and team workflow, rather than on long-running architectural restructuring.

\paragraph{Automated and AI-assisted refactoring.}
Pattern-based and scripted tools for mechanical refactorings have existed for a long time, ranging from IDE refactoring engines to specialized tools such as Coccinelle~\cite{lawall2010finding} and example-based systematic editing frameworks such as LASE~\cite{jacobellis2013lase}.
Despite this, Kim et al.\ report that many developers still perform most refactorings manually and desire more support for refactoring-aware code review, cost/benefit estimation, and automated validation beyond primitive rename or extract operations~\cite{kim2012field}.
More recently, large language models (LLMs) have been proposed as a new vehicle for automated refactoring.
Cordeiro et al.\ empirically evaluated several LLMs on 5{,}194 refactoring commits drawn from 30 open-source Java projects~\cite{cordeiro2026llmrefactoring}.
They showed that production-grade models can perform realistic, commit-level, multi-file refactorings with high test success rates and substantial reductions in code smells, although failures and unintended behavioral changes remain common without strong automated testing.
That work evaluates LLMs offline on curated benchmarks and focuses on code quality metrics and functional correctness.
Bellur et al.\ present CoRenameAgent, a multi-agent, human-in-the-loop tool for coordinated renaming that runs as an IntelliJ IDEA plugin, infers renaming scopes, and propagates rename refactorings across large Java codebases while keeping developers in control through interactive review and transparent status reporting~\cite{bellur2026corenameagent}.
CoRenameAgent shows that integrating agents tightly with IDE refactoring APIs and static program analysis can automate widespread mechanical edits with modest review overhead, but does not examine how such agentic refactorings interact with CI capacity or code review workflows in large, CI-intensive industrial systems, which is our focus in this study.

\paragraph{Developer experience with GenAI tools}
Brandebusemeyer et al.\ conduct a mixed-methods field study of professional developers at SAP to analyze how GitHub Copilot affects developer experience, cognitive load, and interaction patterns in controlled and natural work settings~\cite{brandebusemeyer2026developersexperiencegenerativeai}.
They find that developers are generally satisfied with GenAI, particularly for monotonous, repetitive, and structured tasks, and report perceived efficiency and productivity gains, but that AI interaction can increase cognitive load and that combining multiple interaction types (in-code suggestions and chat) within a single task diminishes efficiency benefits.
He et al.\ report a longitudinal case study of an enterprise ``2$\times$'' AI mandate at a mid-sized B2B software company, analyzing 802 developers and 196{,}212 pull requests over more than two years~\cite{he2026aiwritesfasterhumans}.
They find that per-capita throughput eventually doubles, driven primarily by AI adoption intensity and accumulated use, but that as AI-authored changes reach roughly 90\% of volume, automated review overtakes human review and per-reviewer load roughly doubles while merge and revert rates remain stable on coarse quality metrics.
In contrast to these organization-wide studies, we report on the day-to-day use of an LLM-based command-line coding buddy in a large, closed-source industrial \Cpp{} repository. We study how massive AI-generated remediation interacts with existing CI infrastructure and human review processes, including batching strategies and the social dynamics of getting large numbers of clean-up commits accepted.

\section{Method}

We conducted an exploratory single-case field study of large-scale code cleaning in a large, long-lived industrial software repository.
The purpose is to explore which bottlenecks arise when the programming work of cleaning code becomes very cheap for a single developer assisted by an AI coding buddy.

\subsection{Context}

The exploration took place in a specialized industrial, closed-source repository containing around 10 million lines of code.
The repository is several decades old and has received hundreds of thousands of commits.
Git is used for version control and Gerrit for code review. Commits uploaded in Gerrit need two approved code reviews and must have passed static analysis and verification in the CI environment to be eligible for merging. Each developer has a review dashboard in Gerrit, displaying the status of their own commits and the commits they have been invited to review. It is common practice to invite one's team as well as previous contributors to a file to review a commit.

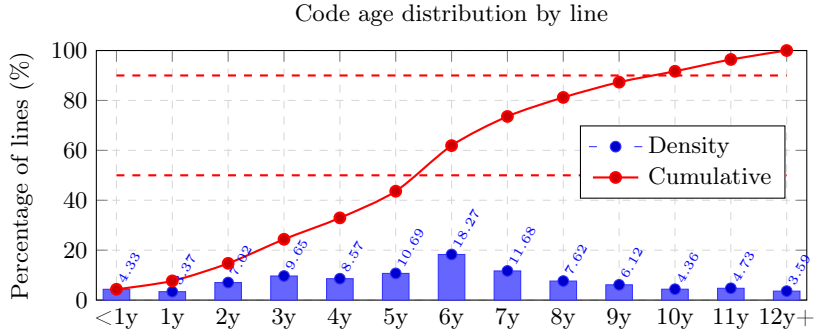
\begin{figure}
    \centering
\begin{tikzpicture}
\begin{axis}[
    width=0.9\textwidth,
    height=0.8\textwidth/2,
    ylabel={Percentage of lines (\%)},
    title={Code age distribution by line},
    symbolic x coords={<1y,1y,2y,3y,4y,5y,6y,7y,8y,9y,10y,11y,12y+},
    xtick=data,
    legend style={at={(0.97,0.7)}, anchor=north east},
    legend cell align={left},
    enlarge x limits=0.03,
    grid=major,
    grid style={dashed, gray!30},
    ymin=0, ymax=100,
]

\addplot+[
    ybar,
    fill=blue!60,
    draw=blue!80,
    nodes near coords,
    nodes near coords style={font=\tiny, rotate=60, anchor=west},
] coordinates {
    (<1y,4.33) (1y,3.37) (2y,7.02) (3y,9.65) (4y,8.57) (5y,10.69)
    (6y,18.27) (7y,11.68) (8y,7.62) (9y,6.12) (10y,4.36) (11y,4.73)
    (12y+,3.59)
};

\addplot+[
    red,
    thick,
    mark=*,
    smooth,
] coordinates {
    (<1y,4.33) (1y,7.70) (2y,14.72) (3y,24.37) (4y,32.94) (5y,43.63)
    (6y,61.90) (7y,73.58) (8y,81.20) (9y,87.32) (10y,91.68)
    (11y,96.41) (12y+,100.00)
};

\addplot[
    red,
    dashed,
    thick,
] coordinates {
    (<1y,50) (12y+,50)
};

\addplot[
    red,
    dashed,
    thick,
] coordinates {
    (<1y,90) (12y+,90)
};

\legend{Density, Cumulative}
\end{axis}
\end{tikzpicture}

\caption{The time since a code line was last touched in the repository. \emph{Takeaway:} It takes five and a half years to half the lines being touched and 10 years to 90\%.}
\label{fig:code_age}
\end{figure}

The age distribution of lines in the repository (\Figure{fig:code_age}) shows that many lines remain untouched for years. Static analysis warnings are typically triggered only on new code or on lines that have been touched by a code change, and the age distribution suggests that it may take a long time before a new static analysis rule has propagated through the source code.

The repository targets several kinds of hardware.
Per commit, the CI chain performs static analysis, runs unit tests with dynamic analysis in a development environment, and builds the software for multiple platforms.
A full commit analysis usually takes 45-90 minutes, and later CI stages run more extensive tests on batches of commits.
In this setting, even small changes are costly to verify, making the interaction between AI-generated remediation and CI capacity particularly salient.

\subsection{Data collection and analysis}

\begin{figure}[tb]
    \centering
\begin{tikzpicture}
    \begin{axis}[
        name=lower,
        width=0.9\textwidth,
        height=0.33\textwidth,
        scale only axis,
        xmin=0, xmax=14,
        ymin=0, ymax=80,
        grid=both,
        grid style={dashed, gray!30},
        every axis/.append style={thick},
        xtick={0,1,2,3,4,5,6,7,8,9,10,11,12,13,14},
        xticklabels={1,2,3,4,5,6,7,8,9,10,11,12,13,14,15},
        x tick label style={font=\scriptsize},
        ylabel={Open commits},
        stack plots=y,
        area style,
        clip=true,
    ]
    \node[draw] at (0.5,60) {1};
    \node[draw] at (4.3,70) {2};
    \node[draw] at (8.9,38) {3};
    \node[draw] at (11.5,67) {4};

    \addplot[fill=orange!50, draw=orange!70!black, line width=0.4pt] coordinates {
        (0,638) (1,638) (2,635) (3,628) (4,26) (5,26) (6,13) (7,13) (8,13)
        (9,12) (10,12) (11,11) (12,11) (13,10) (14,8)
    } \closedcycle;

    \addplot[preaction={fill=green!50}, pattern=north east lines, pattern color=green!70!black, draw=green!70!black, line width=0.4pt] coordinates {
        (0,20) (1,20) (2,20) (3,17) (4,11) (5,11) (6,5) (7,5) (8,5)
        (9,5) (10,5) (11,4) (12,1) (13,1) (14,1)
    } \closedcycle;

    \addplot[fill=violet!50, draw=violet!70!black, line width=0.4pt] coordinates {
        (0,0) (1,0) (2,0) (3,0) (4,0) (5,0) (6,0) (7,0) (8,0)
        (9,40) (10,39) (11,38) (12,34) (13,34) (14,34)
    } \closedcycle;

    \addplot[preaction={fill=cyan!50}, pattern=crosshatch, pattern color=cyan!70!black, draw=cyan!70!black, line width=0.4pt] coordinates {
        (0,1) (1,1) (2,1) (3,1) (4,1) (5,1) (6,1) (7,1) (8,1)
        (9,1) (10,6) (11,7) (12,7) (13,7) (14,8)
    } \closedcycle;

    \addplot[preaction={fill=brown!50}, pattern=north west lines, pattern color=brown!70!black, draw=brown!70!black, line width=0.4pt] coordinates {
        (0,0) (1,0) (2,0) (3,0) (4,0) (5,0) (6,0) (7,0) (8,0)
        (9,0) (10,0) (11,0) (12,7) (13,7) (14,6)
    } \closedcycle;

    \end{axis}

    \begin{axis}[
        name=upper,
        at={(lower.north west)},
        anchor=south west,
        yshift=0.5cm,
        width=0.9\textwidth,
        height=0.2\textwidth,
        scale only axis,
        xmin=0, xmax=14,
        ymin=620, ymax=665,
        grid=both,
        grid style={dashed, gray!30},
        every axis/.append style={thick},
        xtick={0,1,2,3,4,5,6,7,8,9,10,11,12,13,14},
        xticklabels={},
        ytick={620, 640, 660},
        stack plots=y,
        area style,
        clip=true,
        axis x line=box,
        x tick label style={draw=none, font=\empty},
        legend style={
            at={(0.98,0.98)},
            anchor=north east,
            legend columns=2,
            font=\scriptsize,
            draw=none,
            fill=white,
            fill opacity=0.8,
            text opacity=1,
        },
    ]

    \addplot[fill=orange!50, draw=orange!70!black, line width=0.4pt] coordinates {
        (0,638) (1,638) (2,635) (3,628) (4,26) (5,26) (6,13) (7,13) (8,13)
        (9,12) (10,12) (11,11) (12,11) (13,10) (14,8)
    } \closedcycle;
    \addlegendentry{defensive initialization}

    \addplot[preaction={fill=green!50}, pattern=north east lines, pattern color=green!70!black, draw=green!70!black, line width=0.4pt] coordinates {
        (0,20) (1,20) (2,20) (3,17) (4,11) (5,11) (6,5) (7,5) (8,5)
        (9,5) (10,5) (11,4) (12,1) (13,1) (14,1)
    } \closedcycle;
    \addlegendentry{duplicated includes}

    \addplot[fill=violet!50, draw=violet!70!black, line width=0.4pt] coordinates {
        (0,0) (1,0) (2,0) (3,0) (4,0) (5,0) (6,0) (7,0) (8,0)
        (9,40) (10,39) (11,38) (12,34) (13,34) (14,34)
    } \closedcycle;
    \addlegendentry{boolean simplification}

    \addplot[preaction={fill=cyan!50}, pattern=crosshatch, pattern color=cyan!70!black, draw=cyan!70!black, line width=0.4pt] coordinates {
        (0,1) (1,1) (2,1) (3,1) (4,1) (5,1) (6,1) (7,1) (8,1)
        (9,1) (10,6) (11,7) (12,7) (13,7) (14,8)
    } \closedcycle;
    \addlegendentry{potential races}

    \addplot[preaction={fill=brown!50}, pattern=north west lines, pattern color=brown!70!black, draw=brown!70!black, line width=0.4pt] coordinates {
        (0,0) (1,0) (2,0) (3,0) (4,0) (5,0) (6,0) (7,0) (8,0)
        (9,0) (10,0) (11,0) (12,7) (13,7) (14,6)
    } \closedcycle;
    \addlegendentry{class hierarchy}

    \end{axis}


    %
    \begin{axis}[
        name=helper,
        at={(lower.south west)},
        anchor=south west,
        width=0.9\textwidth,
        height=0.33\textwidth,
        scale only axis,
        xmin=0, xmax=14,
        ymin=0, ymax=80,
        hide axis,
        no markers,
    ]
        \coordinate (ld0) at (axis cs:0,80);
        \coordinate (ld391) at (axis cs:3.91,80);
    \end{axis}

    \begin{axis}[
        name=helper2,
        at={(lower.north west)},
        anchor=south west,
        yshift=0.5cm,
        width=0.9\textwidth,
        height=0.2\textwidth,
        scale only axis,
        xmin=0, xmax=14,
        ymin=620, ymax=665,
        hide axis,
        no markers,
    ]
        \coordinate (ud0) at (axis cs:0,620);
        \coordinate (ud3015) at (axis cs:3.013,620);
    \end{axis}

    \fill[orange!50, draw=orange!70!black, line width=0.4pt]
        (ld0) -- (ld391) -- (ud3015) -- (ud0) -- cycle;

    \coordinate (left-bot) at (lower.north west);
    \coordinate (left-top) at ([yshift=0.5cm]lower.north west);
    \draw[thick] (left-bot)
        -- ([xshift=3pt]$0.2*(left-top) + 0.8*(left-bot)$)
        -- ([xshift=-3pt]$0.4*(left-top) + 0.6*(left-bot)$)
        -- ([xshift=3pt]$0.6*(left-top) + 0.4*(left-bot)$)
        -- ([xshift=-3pt]$0.8*(left-top) + 0.2*(left-bot)$)
        -- (left-top);

    \coordinate (right-bot) at (lower.north east);
    \coordinate (right-top) at ([yshift=0.5cm]lower.north east);
    \draw[thick] (right-bot)
        -- ([xshift=3pt]$0.2*(right-top) + 0.8*(right-bot)$)
        -- ([xshift=-3pt]$0.4*(right-top) + 0.6*(right-bot)$)
        -- ([xshift=3pt]$0.6*(right-top) + 0.4*(right-bot)$)
        -- ([xshift=-3pt]$0.8*(right-top) + 0.2*(right-bot)$)
        -- (right-top);


\end{tikzpicture}

\vspace{0.1cm}
\begin{tikzpicture}
    \begin{axis}[
        width=0.9\textwidth,
        height=0.17\textwidth,
        scale only axis,
        xmin=0, xmax=14,
        ymin=0, ymax=20,
        grid=both,
        grid style={dashed, gray!30},
        every axis/.append style={thick},
        xtick={0,1,2,3,4,5,6,7,8,9,10,11,12,13,14},
        xticklabels={1,2,3,4,5,6,7,8,9,10,11,12,13,14,15},
        x tick label style={font=\scriptsize},
        ylabel={Commits per day},
        legend style={
            at={(0.98,0.98)},
            anchor=north east,
            legend columns=1,
            font=\scriptsize,
            draw=none,
            fill=white,
            fill opacity=0.8,
            text opacity=1,
        },
    ]

    \node[draw] at (2.8,12) {5};
    \node[draw] at (7.5,3) {6};

    \addplot[color=red, mark=triangle, line width=1.2pt] coordinates {
        (0,0) (1,3) (2,15) (3,0) (4,10) (5,8) (6,2) (7,0) (8,1) (9,2) (10,6) (11,7) (12,1) (13,2) (14,4)
    };
    \addlegendentry{Review nudges}

    \addplot[color=teal, mark=diamond, line width=1.2pt, dashed] coordinates {
        (0,0) (1,2) (2,10) (3,9) (4,0) (5,19) (6,0) (7,0) (8,1) (9,1) (10,3) (11,7) (12,0) (13,2) (14,4)
    };
    \addlegendentry{Merged}

    \end{axis}
\end{tikzpicture}

    \caption{\footnotesize Open commits by category (top) and daily review requests and merges (bottom) over 15 working days. The y-axis of the upper chart is broken to show both the initial peak and the subsequent detail; the sharp drop on day~5 reflects the mass abandonment of struct init commits. The lower chart shows the daily numbers of requests for review and merged commits.}
    \label{fig:open_commits}
\end{figure}
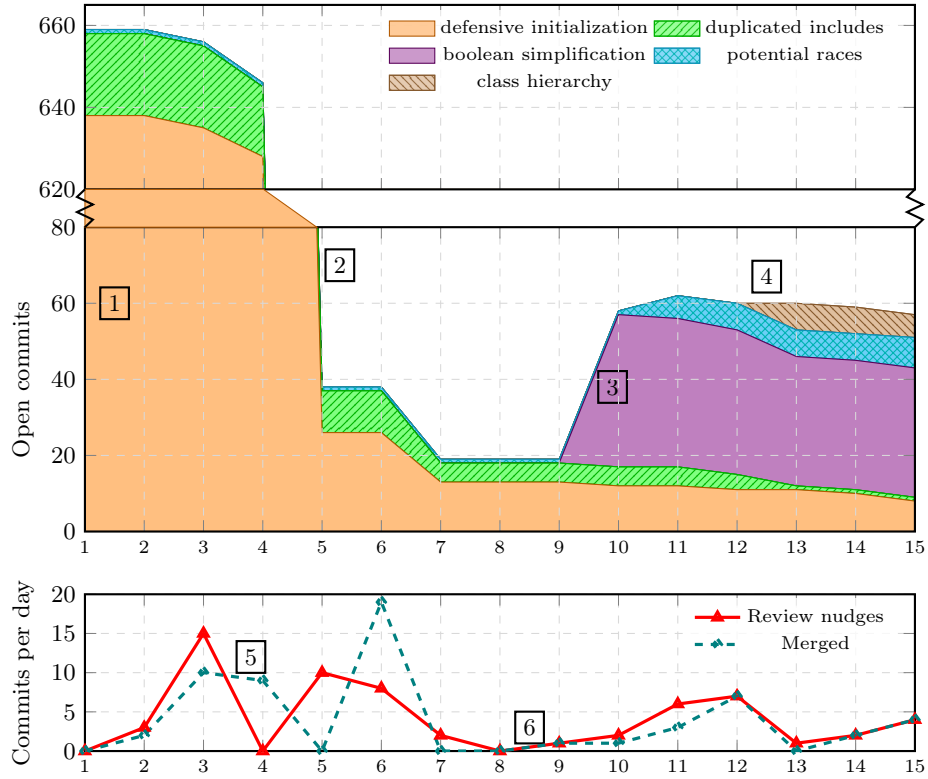

The exploration was carried out over 15 working days (May 19 – June 8, 2026) by one experienced developer using the AI coding buddy to identify and remediate code degradation issues. Issues were chosen by recommendation from the coding buddy, by prior experience from the developer and by request from other team members.
We collected three complementary data sources:

  \paragraph{\bf Diary.} The exploring developer kept a day-by-day diary documenting planned activities, interactions with the AI assistant, perceived bottlenecks, and notable incidents (e.g., CI overloads, review discussions).
  \paragraph{\bf Team chat.} We collected anonymized chat logs between the developer and their team from the main communication channel during the same period, capturing review requests, reactions, and coordination around the AI-gene\-rated changes.
  \paragraph{\bf Gerrit metadata.} Gerrit metadata. From Gerrit, we extracted quantitative data for all changes associated with the exploration using the Gerrit SSH API\footnote{Example query: \texttt{ssh <host> gerrit query -{}-format=JSON -{}-current-patch-set -{}-files "owner:<username> after:2026-05-18 message:"<identifier>"}, with pagination via \texttt{-{}-start} for large result sets.}. Since Gerrit's built-in date filters (\texttt{after:}/\texttt{before:}) operate on the \texttt{lastUpdated} field rather than creation time, we queried broadly and then post-filtered results in Python by the \texttt{createdOn} timestamp to identify commits created within the 15-day window. For each commit, we recorded its status (merged, abandoned, or open), creation date, and close date (derived from \texttt{lastUpdated} for merged or abandoned commits). Per-file insertions and deletions were extracted from the current patch set, excluding the synthetic \texttt{/COMMIT\_MSG} pseudo-file. Line counts in \Table{tab:remediation-overview} represent insertions plus the absolute value of deletions. Open-commit counts in \Figure{fig:open_commits} are computed as the number of commits created on or before a given working day that had not yet been merged or abandoned by the end of that day.

\paragraph{} We linked these sources using timestamps and Gerrit change identifiers, enabling us to trace each remediation activity across AI interaction, developer intent, team communication, and CI outcomes. 
We used Gerrit metadata to compute descriptive statistics (e.g., numbers and sizes of commits per remediation category, merges, abandonments, and CI failure modes). 
For the diary and chat logs, we conducted an inductive qualitative analysis using Taguette: one author performed initial open coding independently from the exploring author, followed by collaborative grouping and refinement of codes into themes related to CI, code review, and team workflow. All personal data were anonymized before analysis.

\section{Results}
\label{sec:results}

In this section, we address \RQ{1} by quantifying how AI-assisted remediation activities interacted with code review and continuous integration (CI), and we address \RQ{2} by describing which socio-technical bottlenecks we encountered when source editing became cheap.

\subsection{Scale and progression of AI-assisted remediation}
\label{scale}

Figure~\ref{fig:open_commits} shows how the number of open commits,
review requests, and merged changes evolved over the 15 working days, broken
down by remediation category (duplicated \texttt{\#include} directives, struct
initialization, redundant boolean comparisons, data race fixes, and a module
refactoring).

We started the exploration by asking the AI coding buddy to propose remediation
targets from the commit history. The buddy suggested three initial candidates:
duplicated \texttt{\#include}-statements, a specific race condition, and name
``stuttering'' (for example, \texttt{ECB\_ECB\_FAULT} rather than
\texttt{ECB\_FAULT}). The developer first addressed duplicated includes and
uploaded 20 directory-based commits to Gerrit. 

Next, the developer drew on prior knowledge of the codebase and decided to
tackle uninitialized primitives in structs. To keep reviews simple, the developer
initially instructed the buddy to upload one commit per file, which produced 639
commits. Event~\fbox{1} in
Figure~\ref{fig:open_commits} marks this initial wave of duplicate-\texttt{\#include}
and struct-init clean-ups. CI maintainers soon contacted the developer and pointed out that these
per-file uploads had saturated the build-on-commit CI chain; they purged the
queue and manually retriggered jobs when appropriate. 

After this incident, the developer realized that hundreds of tiny commits would
both consume excessive CI capacity and take too long to attract reviewer
attention. On day~5, the developer therefore asked the buddy to abandon the open
per-file struct commits and re-upload the same changes batched per directory.
This step replaced 639 per-file commits with 24 larger, directory-based commits.
Event~\fbox{2} in Figure~\ref{fig:open_commits} corresponds to this switch in
granularity and the sharp drop in open struct-initialization commits.

Reviewer feedback then led us to tune commit size. Some early batched uploads
touched more than 100 files, and one reviewer reported: \emph{``I opened every file and
look (sic) into it, less than 5 seconds for each file, and it still takes a long
time.''}. In response, the developer reasoned that \emph{``a trivial review shouldn’t
take more than 5 minutes''} and concluded that they \emph{``shouldn’t post more than 60
files per commit''}. We adopted this cap and aimed for commits that reviewers
could process in about five minutes.

Once the duplicate-\texttt{\#include} and struct-init clean-ups were under control, the
developer instructed the buddy to find explicit boolean comparisons (for example,
\texttt{a == true} rather than \texttt{a}, or \texttt{b == false} rather than
\texttt{!b}) and to group fixes so that no commit contained more than 60 files,
with a 15-minute delay between uploads. The buddy identified many such
instances and produced 56 commits in a single day under these batching rules.
Event~\fbox{3} in Figure~\ref{fig:open_commits} corresponds to this high-cadence
boolean-comparison clean-up.

The buddy had already fixed one possible race condition on day~1 (a pattern it identified during the initial commit-history scan). Later, colleagues asked the developer to address a related concurrency pattern involving unprotected shared variables. The buddy found several instances and fixed them mechanically, producing a small series of further race-related remediation commits in the third week.

On day~12, a meeting about the repository structure prompted the developer to
try the buddy on a larger design task. The buddy analyzed a class hierarchy,
identified object-oriented design optimization opportunities, and proposed changes to simplify
the API and reduce redundancy. The resulting 13 commits depended on each other
and required carefully ordered merging, unlike the earlier mostly independent
mechanical clean-ups. 

In the final days of the exploration, the developer and buddy shifted focus to
fixing CodeChecker findings and CI build failures across multiple platforms. In
one notable case, a struct-init change had to be reverted because it
broke assumptions in legacy builds; the developer and buddy redesigned the
change to use a dedicated ``undefined'' marker value and re-applied the safe
initializations on top of the revert. Event~\fbox{4} in
Figure~\ref{fig:open_commits} represents this tail of CI and
static-analysis remediation and the need to iteratively redesign seemingly
mechanical fixes.

Table~\ref{tab:remediation-overview} summarizes these AI-assisted remediation activities: 20 duplicate-\texttt{\#include} commits, 639 per-file \texttt{struct}-init commits followed by 24 directory-based \texttt{struct} commits, 56 redundant boolean-comparison commits, 8 potential data-race fixes, and 13 module-refactoring commits. Together, these changes touched thousands of lines of code and made CI capacity, reviewer attention, and change orchestration first-order constraints. During the exploration, the developer kept a commit cadence an order of magnitude higher than the median commit cadence per developer.


\begin{table}[t]
\scriptsize
  \centering
  \caption{Overview of AI-assisted remediation activities during the exploration. \emph{Takeaway:} Trivial clean-ups can become massive changes.}
  \label{tab:remediation-overview}
  \setlength{\tabcolsep}{3pt}
  \begin{tabular}{lrrrl}
    \toprule
    \tiny Category & \tiny Commits & \tiny Files & \tiny Lines & \tiny Brief description \\
    \midrule
    Duplicated \texttt{\#include}s & 20 & 207 & 244 & Remove redundant include directives \\
    Struct initialization (per-file) & 639 & 639 & 81,160 & Give deterministic defaults to primitives \\
    Struct initialization (batched) & 24 & 813 & 82,990 & Directory-based batched commits \\
    Redundant bool comparisons & 56 & 865 & 9,096 & Replace \lstinline{== true}/\lstinline{== false} \\
    Data race fixes & 8 & 16 & 351 & Fix instances of data-race patterns \\
    Module refactoring & 13 & 124 & 21,130 & Class hierarchy and API refactoring \\
    \bottomrule
  \end{tabular}
\end{table}

\subsection{Qualitative dynamics from the codebook}

The codebook aggregates
26 highlights tagged \emph{nudging-for-review}, 12 tagged
\emph{describing-agentic-process}, 18 tagged \emph{experiencing-problem}, 8 tagged
\emph{closing-changesets}, and 6 tagged \emph{state:delivery-stop}, along with
smaller but important categories such as \emph{overloading-ci},
\emph{discussing-changeset-size}, \emph{considering-process-changes},
\emph{planning-next-steps}, and \emph{feeding-error-msg-to-agent}.

The \emph{nudging-for-review} tag dominates both chat and diary material, and it
explains why the merged-commit curve in Figure~\ref{fig:open_commits} closely
tracks explicit review requests (see \fbox{5} and \fbox{6}). The developer repeatedly posts messages such as
\emph{``Low risk, verified, has +1; please review''} and \emph{``Thank you for helping me get
a nice merge-streak!''} and lists \emph{``low-hanging fruit''} and \emph{``very simple''}
commits. These messages show that the developer could not rely on colleagues discovering
changes in Gerrit but instead had to actively solicit reviews to keep the stream of
changes moving.

Tags such as \emph{\small discussing-changeset-size}, \emph{\small receiving-response:time-spent},
and \emph{\small asking-about-process} capture explicit negotiation of suitable commit
granularity. One team member asks: \emph{``would it be better to combine them into
10--20 commits?''} and suggests that \emph{``50--100 files sound good''}. The developer
replies that they \emph{``still try to strike a balance between overloading the build
system and overloading the reviewers''} and proposes the 60-file commit cap
mentioned above. These exchanges directly underpin events~\fbox{2} and \fbox{3} in
Figure~\ref{fig:open_commits} and show how we tuned commit size to match human
review capacity.

The \emph{\small overloading-ci} and \emph{\small experiencing-problem} tags highlight how
quickly naive batching can overwhelm CI. In the diary, the developer writes:
\emph{``Inadvertently, I had flooded the build queue with 639 commits''} and describes
the incident as an \emph{``attack''} on the CI chain. They note that they \emph{``spent most
of the day apologizing to DevOps''} and later report that many duplicate-\texttt{\#include}
commits remain in \emph{``aborted verification''} state until they rebase and retrigger.
These episodes align with event~\fbox{2} in
Figure~\ref{fig:open_commits} and show that, once mechanical editing becomes
cheap, CI infrastructure quickly becomes a hard bottleneck that developers must
actively manage.

Emotional and organizational impacts appear in
\emph{\small feeling-some-anxiety} and \emph{\small state: delivery-stop}. The developer writes
that they feel \emph{``a little bit of stomach ache''} when delivery stops coincide with
large-scale initialization changes and repeatedly notes days where a delivery
stop prevents merging even though many commits are ready. These states correspond
to flat segments in the merged-commit curve in
Figure~\ref{fig:open_commits} and illustrate how release freezes interact with
AI-assisted remediation.

Several tags, including \emph{\small describing-agentic-process},
\emph{feeding-error-msg-to-agent}, and \emph{\small verifying-changesets-locally}, show
how the developer gradually integrates the buddy into the toolchain. The
developer describes how \emph{``The agent went to work, and I went to lunch''} and later
reports that they copy-paste CI error messages so that the buddy can \emph{``correctly
identify the root cause''} and amend failing commits. They instruct the buddy to
always run available static analysis before uploading, to update commits by creating new patch
sets rather than abandoning and recreating them, and to locally verify and amend
26 failing commits. These interactions correspond mainly to event~\fbox{3} in
Figure~\ref{fig:open_commits} and show that we can use AI agents to absorb
menial follow-up work, but only if we give them structured access to build and
analysis information.

Finally, tags such as \emph{\small closing-changesets}, \emph{\small planning-next-steps}, and
\emph{\small spreading-info-in-org} reveal that the experiment is both technical and
social. The developer notes days where they \emph{``merged 7 initialization-commits
and 3 dup include today''} and where they prepare a presentation for an internal
AI group, stating that this \emph{``makes me take stock of the experiment so far''}. They
also record plans for future bulk improvements, such as addressing redundant constructors, and missing \texttt{!=} operators. New ideas for improvement
occur throughout the timeline and help explain why the curves in
Figure~\ref{fig:open_commits} show sustained activity rather than a short-lived
spike.

Overall, the quantitative evolution in Figure~\ref{fig:open_commits} and the
qualitative patterns in the codebook jointly indicate
that, once mechanical editing becomes cheap through AI assistance, CI capacity,
review effort, and explicit coordination around batching, commit granularity, and
tool integration become primary constraints. The coded episodes from the diary and team chat
highlight how developers and teams renegotiate processes, tooling, and social
practices to make massive AI-assisted remediation sustainable in a large
industrial repository.
\section{Threats to Validity}
  
\paragraph{Credibility.} Our study relies on a single developer's diary, team chat, and Gerrit metadata over 15 working days. The developer is also a co-author, which means the diary is not an independent observation: it reflects the practitioner's own sense-making. We mitigate this by triangulating diary claims against Gerrit timestamps and team chat, and by having a separate author perform the initial qualitative coding. However, we cannot rule out selective reporting in the diary or that the developer unconsciously framed events to fit an emerging narrative.

\paragraph{Analyzability.} The qualitative coding was performed inductively by one author and then refined collaboratively. Since the exploring developer is a co-author, there is a risk of insider bias in interpreting diary and chat excerpts. We addressed this by having the initial coding done independently from the exploring author, but a second independent coder would have strengthened inter-rater reliability. Quantitative claims (commit counts, file counts, line counts) are derived directly from Gerrit metadata and are reproducible, though the filtering criteria (e.g., excluding \texttt{COMMIT\_MSG} pseudo-files, the 15-day window boundary) involve judgment calls that we document explicitly.
  
\paragraph{Transparency.} We report the exact Gerrit queries, date boundaries, and counting rules used to produce \Table{tab:remediation-overview} and \Figure{fig:open_commits}. The repository is closed-source, which prevents full replication, but the methodology (query Gerrit, filter by creation date, compute open-at-end-of-day counts) is described in sufficient detail to be applied in any Gerrit-based environment.
  
\paragraph{Usefulness.} This is a single-case exploratory study of one developer using one AI coding buddy in one large \Cpp{} repository over fifteen working days. Our findings are not statistically generalizable. However, the phenomena we observe — CI saturation from high-cadence commits, review bottlenecks from bulk mechanical changes, and the need to negotiate commit granularity — are common in industrial environments with build-on-commit CI, static analysis, and large legacy systems. Systems with smaller codebases, lighter-weight CI, different language ecosystems, or substantially different team structures may experience other bottlenecks. Developer-centered mixed-methods studies of AI tools in industrial settings report related tensions between productivity, cognitive load, and review capacity \cite{brandebusemeyer2026developersexperiencegenerativeai,he2026aiwritesfasterhumans}, suggesting these bottlenecks are likely to recur as AI-assisted development becomes more prevalent.
\section{Discussion}
\label{sec:discussion}

Cleaning code in large, old repositories quickly leads to massive code changes, even when each individual edit is small and mechanical.
We discuss implications for technical debt remediation, code review, continuous integration (CI), and future change-orchestration models.

As indicated by the age distribution of lines in the repository (\Figure{fig:code_age}), it may take a long time for change-based static analysis to propagate opportunistically through the codebase. In practice, code smells tend to accumulate over time.
Combined with the static-analysis triggers observed in our exploration, this suggests that many smells are better treated as \emph{systematic} remediation targets at scale rather than as purely opportunistic clean-ups.

\paragraph{Bulk changes become massive.}
Systematic code cleaning efforts very quickly become massive code changes.
Several smell categories were so widespread that addressing them comprehensively touched thousands of lines across hundreds of files.
At the same time, many edits were functionally independent: each duplicated \texttt{\#include} could be fixed in isolation; each uninitialized primitive could be given a deterministic default independently of the others.
A naive strategy of creating one-line commits for each fix would have choked the build-on-commit CI chain and generated an unmanageable number of review requests.
Conversely, a few ``big-bang'' commits would have been tedious to review, fragile with respect to merge conflicts, and hard to revert selectively.
Large-scale remediation therefore requires deliberate control over the granularity of both commits and review units: commits must be small enough to reason about, but large enough that CI and human reviewers are not overwhelmed by their number.
Agentic refactoring tools such as CoRenameAgent already treat coordinated renamings as semantic change sets with explicit scopes and refactoring reports in an IDE~\cite{bellur2026corenameagent}.
Our findings suggest that similar abstractions at the repository level---for example change sets corresponding to ``fix all instances of warning X''---are needed to keep bulk mechanical remediation comprehensible and reviewable when it touches thousands of lines across hundreds of files.

\paragraph{Continuous integration infrastructure is a bottleneck.}
Build-on-commit conventions may not scale when commit cadence increases due to mechanical batch changes generated by an AI assistant.
Predictive commit skipping~\cite{abdalkareem2019commits,mhalla2024detecting} could, in principle, reduce CI load for repetitive, low-risk clean-ups, and batching multiple commits per build~\cite{9392370} can increase utilization of available resources.
However, our three-week experience in a complex system with many build variants and several static-analysis tools suggests that accurately predicting ``safe'' commits is difficult: the risk of missing a change that triggers a failure later in the CI chain is high.
Effective remediation at machine speed exposes CI infrastructure as a first-order constraint that must be explicitly managed, for example, by dedicating CI batches to mechanical clean-ups and tuning latency expectations separately from feature work.
Similar dynamics appear in longitudinal studies of enterprise AI mandates, where per-capita throughput eventually doubles but review and CI capacity become the limiting factors and organizations turn to automated review and batching to absorb the added volume~\cite{he2026aiwritesfasterhumans}.

\paragraph{Tool integration as an enabler.}
A recurring bottleneck in our exploration was the limited integration between the AI coding buddy and the surrounding toolchain.
The coding buddy could access Gerrit and some static-analysis tools, but not build-system logs or all analysis outputs, and it was not connected to issue-handling systems.
As a result, the developer had to manually copy-paste CI and CodeChecker output into the chat and orchestrate follow-up changes by hand.
This lack of end-to-end integration made remediation more tedious, prolonged the tail of follow-up fixes, and reduced the potential benefits of automation.
We hypothesize that making AI coding assistants first-class citizens in the toolchain, with direct, structured access to build systems, static analysis, and issue trackers, and with appropriate safeguards, would be essential for scaling systematic remediation beyond single-developer explorations.
In contrast, IDE-integrated agentic frameworks such as CoRenameAgent demonstrate how multi-agent systems can use refactoring APIs and static program analysis to automate coordinated mechanical edits with modest interaction overhead~\cite{bellur2026corenameagent}; bringing similar integration to CI and analysis pipelines appears necessary to fully realize the potential of AI-assisted remediation in large industrial repositories.

\paragraph{Review at scale and the role of automation.}
Human review of every individual clean-up commit is unlikely to be feasible when development cadence increases.
Bacchelli and Bird find that the main drivers for code review among developers are (a) finding defects, (b) code improvement, (c) alternative solutions, (d) knowledge transfer, and (e) team awareness~\cite{bacchelli2013expectations}.
Monperrus argues that all of these goals can, in principle, be served at least as well by automated review agents~\cite{monperrus2026end}.
This is particularly compelling for the bulk cleaning tasks we focused on: tests and static analysis are better suited to finding defects in mechanical refactorings; the code improvement is largely defined by the remediation pattern; there is little room for meaningful alternative solutions; and knowledge transfer and team awareness appear to saturate quickly, often after reviewing the first few representative commits.
Beyond that point, additional reviews of the same pattern offer diminishing returns.

Our results suggest that, especially for bulk remediation of static-analysis issues, the developer, reviewers, and the CI system do not necessarily need to operate on the same unit of change.
In many cases, each individual line-level fix is atomic in the sense that it could, in principle, be committed, reviewed, built, and merged independently.
Practically, however, it is more effective to separate these concerns: commits can be generated in bulk by the AI assistant, reviewers can sample and review representative subsets or aggregated diffs per remediation pattern, and CI builds can be executed on carefully chosen batches of changes. If commits touch disjunct sets of files, they might be possible to build together yet still allow tracing of errors back to the offending commit.
Designing such decoupled workflows, where commit granularity, review granularity, and build granularity are tuned independently, appears essential for making systematic, AI-assisted code cleaning sustainable in very large industrial repositories.
This perspective aligns with evidence from enterprise-wide AI rollouts, where AI-authored changes increasingly ship under automated review and human review load shifts toward approving representative samples and higher-risk changes rather than inspecting every individual mechanical edit~\cite{he2026aiwritesfasterhumans}.
It also resonates with developer-centered studies of GenAI tools that report increased cognitive load from validating and adapting AI outputs, alongside perceived productivity gains for monotonous, repetitive tasks~\cite{brandebusemeyer2026developersexperiencegenerativeai}.

\paragraph{Implications for change orchestration and future work.}
Our experience highlights a more fundamental limitation of current workflows: the same unit of change (typically a commit or change list) is reused as the development atom, the review atom, the CI/integration atom, and the history atom.
For massive, mostly independent, AI-generated clean-ups, these concerns need not coincide.
Conceptually, one can instead treat a semantic \emph{change set}, for example ``fix all instances of warning X'', as the primary object and derive different projections of that change set for different consumers: reviewer-sized slices for human review, CI batches chosen for test efficiency and failure isolation, fine-grained integration steps (down to single-line edits) for safe merging and rollback, and history groupings optimized for later comprehension.
Practically, this implies that organizations should
\begin{enumerate*}[label=(\roman*)]
    \item explicitly represent bulk remediation work as semantic change sets rather than ad-hoc clouds of commits,
    \item tune commit size and composition to human review latency targets, and
    \item decouple review granularity from CI granularity when batching AI-generated edits.
\end{enumerate*}

Designing and evaluating such a change-orchestration layer, sitting on top of existing version-control systems and treating commits as an output rather than the fundamental unit of work, is a natural direction for future research.
Agentic refactoring tools and enterprise AI rollouts already provide early examples of semantic change sets and aggregated views over large numbers of AI-generated edits~\cite{bellur2026corenameagent,he2026aiwritesfasterhumans}; extending these ideas to CI and review orchestration, and empirically validating their impact on throughput, failure modes, and reviewer load, is a promising avenue for future work.

\section{Conclusions and Implications}

\paragraph{Answer to \RQ{1}:}
We asked how massive AI-assisted code remediation affects CI, code review practices, and team coordination in a large industrial repository.
In our three-week exploration, high-cadence, AI-generated clean-ups saturated the build-on-commit CI chain, even when individual edits were small. This saturation required renegotiating CI usage with maintainers and switching from per-file commits to directory-based batches and caps on the number of files per change.
The resulting stream of changes also stressed review capacity: hundreds of small commits proved impractical to review individually, motivating tuned commit granularity and explicit review requests via the team chat to obtain timely feedback and merges.
Overall, when mechanical editing becomes cheap through AI assistance, CI capacity, review effort, and explicit coordination around batching and review solicitation become first-order constraints that shape how AI-assisted remediation can be integrated into everyday development.
Field studies of GenAI interaction similarly report that developers experience both productivity gains and increased cognitive load when using AI tools, and that interaction choices and workflow design strongly influence whether AI support is helpful or burdensome~\cite{brandebusemeyer2026developersexperiencegenerativeai}.

\paragraph{Answer to \RQ{2}:}
We asked which socio-technical bottlenecks constrain massive code remediation when source editing becomes cheap through AI assistance.
Our exploration shows that CI capacity, reviewer attention, and change orchestration become primary constraints: naive per-file commits overload build-on-commit CI and generate unmanageable review queues, whereas carefully batched commits and limits on files per change are needed to sustain AI-assisted remediation at scale.
Similar bottlenecks have been observed in organization-wide AI rollouts, where throughput gains from AI tools are accompanied by increased review load and a shift toward automated review~\cite{he2026aiwritesfasterhumans}.

\paragraph{For practitioners}
\begin{enumerate*}[label={\roman*)}]
    \item Represent bulk remediation as semantic change sets, not
  ad-hoc commits.
  \item Cap commit size to human focus capacity ($\sim 60$ files, $\sim 5$ minutes).
  \item Decouple CI batching from review granularity.
  \item Give AI assistants structured access to CI toolchain systems, such as build logs and static analysis.
  \item Expect to actively solicit reviews; changes may not attract attention on their own.
\end{enumerate*}

\paragraph{For the research community.} 
\begin{enumerate*}[label={\roman*)}]
    \item Design and evaluate change-orchestration layers that
  treat commits as derived artifacts rather than fundamental units.
  \item Investigate predictive CI skipping and commit merging for repetitive mechanical clean-ups.
  \item Study how automated review can absorb the review load from bulk remediation.
  \item Conduct longitudinal studies of AI-assisted remediation beyond single-developer explorations.
\end{enumerate*}

\begin{credits}\small
\paragraph{\bf Data Availability} The study is conducted in situ in an industrial context. Raw data may contain sensitive information and is not shared.
{\paragraph{\bf \ackname}\small This work was partially supported by the Wallenberg AI, Autonomous Systems and Software Program (WASP) funded by the Knut and Alice Wallenberg Foundation, and the Competence Centre {\sc NextG2Com} funded by the VINNOVA program for Advanced Digitalisation with grant number 2023-00541. We thank Kristian Wiklund and Leif Jonsson for their constructive review comments on the paper.
\paragraph{\bf \discintname} The authors employed by Ericsson
AB report their affiliation; the authors declare no other competing interests.}
\end{credits}
%
%
%
\bibliographystyle{splncs04}
\bibliography{refs}

\end{document}